%% file: csj_periodic150.tex
\documentclass[usenatbib]{mnras}
\usepackage{epsfig}
\usepackage{amsmath,amssymb}
\usepackage{xcolor}
\usepackage{booktabs}
\usepackage{pdflscape}
\usepackage{textcomp}

\usepackage{epsfig}
\usepackage{soul}
\usepackage{aas_macros}  
\usepackage{chngcntr}
\graphicspath{{figures/}}

\input{mydefs}

\usepackage{graphicx}

\title{Merged white dwarfs and nucleosynthesis}

\author[C.S.~Jeffery \& X.~Zhang]{C. Simon Jeffery$^{1}$\thanks{email: simon.jeffery@armagh.ac.uk} and X. Zhang$^{2}$\thanks{email: zxf@bnu.edu.cn}\\
$^{1}$Armagh Observatory and Planetarium, College Hill, Armagh BT61 9DG, UK\\
$^{2}$Department of Astronomy, Beijing Normal University, Beijing, 100875, People's Republic of China }

\begin{document}
\date{Accepted \ldots. Received \ldots; in original form \ldots}
\pagerange{\pageref{firstpage}--\pageref{lastpage}} \pubyear{2021}
\maketitle
\label{firstpage}

\begin{abstract}
Orbital decay mechanisms argue that double white dwarf mergers are inevitable, but  extremely rare. 
Whilst some mergers result in explosions, the survivors re-ignite helium and burn brightly for tens of thousands or millions of years. 
Candidate survivors include extreme helium stars, R\,CrB variables and various classes of helium-rich subluminous star. 
Nuclear waste on the survivors' surfaces provides evidence of the stars' nuclear history prior to and their nucleosynthesis during the merger. 
Extensive and deep spectroscopic surveys offer rich prospects for future discoveries. 
\end{abstract}





\section{Introduction}

There is overwhelming evidence that many classes of binary star pass through a phase in which the components coalesce to form a single star. 
Prerequisite to coalescence is an orbital decay mechanism, including  any of gravitational-wave (GW) radiation,  magnetic-wind braking (MWB),  mass transfer through Roche-lobe overflow (RLOF) or common-envelope ejection (CEE), or external hardening by the Kozai mechanism. 
Susceptible binaries include stars at nearly every stage of evolution, including binary main-sequence stars (e.g. W\,UMa systems and massive binaries: MWB, RLOF), red giants (RLOF, CEE), and compact stars (e.g. double white dwarfs, double neutron stars, double black holes: MWB, RLOF, GWR).
Evidence comes from observations of orbital decay ($\dot{P}$: PSR 1913+16, J0651+2844) \citep{psr1913+16,hermes12b}, from unusual classes of novae (e.g. V838\, Mon, V1309\,Sco) \citep{soker06,v1309sco,ivanova13}, from GW bursts and other explosive events (e.g. GW 170817 = GRB 170817A) \citep{gw170817,grb170817}, and from single stars that can only be understood in the context of post-merger evolution (e.g. blue stragglers, R\,CrB stars, helium-rich subdwarfs) \citep{leonard89,saio02,zhang12a}. 
In terms of the latter, the surface chemistry of stars that have coalesced provides a unique perspective on the nuclear structure of the progenitor stars, and of the physics of the merger event. 
Each atomic (or nuclear) species visible on the surface of a post-merger remnant tells a different part of the story.

This paper outlines the physics of double white dwarf mergers and what atoms have told us about them in the last century. 
\S\,2 describes the formation of double white dwarf binary star systems and the conditions for a merger to occur.
\S\,3 introduces the physics of the merger itself.
\S\,4 discusses some of the possible merger outcomes, concentrating on stars that survive rather than explosions.
\S\S\,5 and 6 explore how observations of merger survivors can provide tests of theory for two classes of double white dwarf merger models.  

\begin{figure*}
\centering\includegraphics[trim=30 30 00 00, clip, height=.40\textheight]{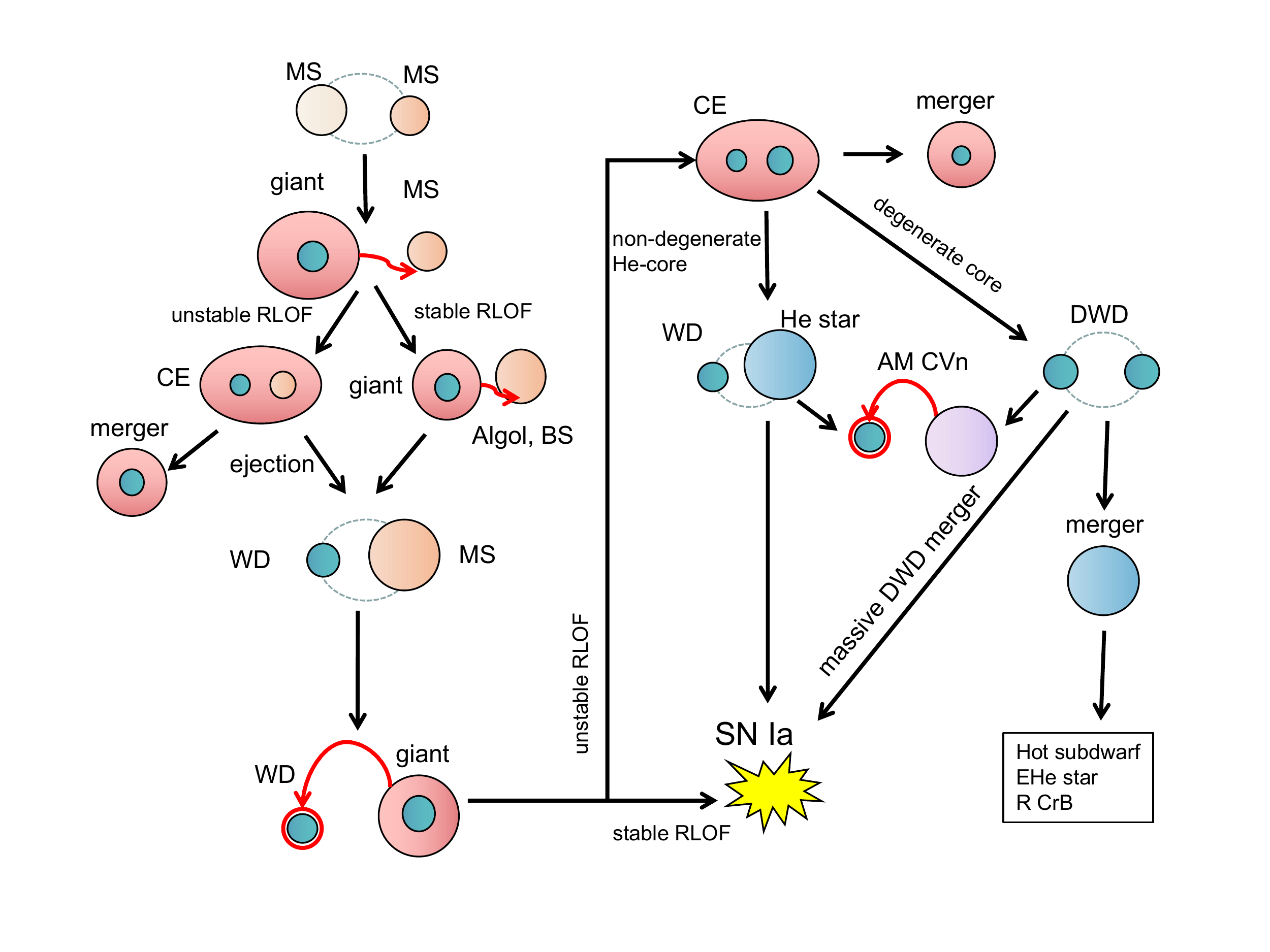}
\caption{A simplified schematic of possible channels leading to the formation of double white dwarf binaries and their subsequent interactions. Abbreviations used include MS: main-sequence stars, RLOF: Roche-lobe overflow, CE: common envleope, BS: blue straggler, WD: white dwarf, He: helium, AM CVn: an interacting double white dwarf binary, SN\,Ia: a Type Ia supernova. }
\label{fig_channels}
\end{figure*}

\section{Double white-dwarf mergers}

\subsection{DWD formation}

White dwarfs (WD) represent the final state of most stars, the remnant left after all nuclear fuel and gravitational potential has been spent, and  $\gtrsim8\%$ of all stars in the Solar neighbourhood. 
Whilst $\sim0.6\Msolar$ carbon-oxygen white dwarfs are the usual end-state of single stars with mass $0.5 - 8\Msolar$, observed WD have masses from $0.1 - 1\,\Msolar$, a range which can only be explained as a consequence of interactions between stars in binary or multiple star systems. 

The essential physics requires two stars to be in an orbit sufficiently compact that, when one star expands to become a red giant, material spills outside the gravitational volume (Roche lobe) where it can be identified with its parent. 
This mass may fall onto its companion or be ejected from the binary system altogether. 
Depending on initial conditions and other physics, the binary orbit can shrink or widen, mass transfer can continue quietly or accelerate, and both stars can radically alter their structure and commence a new phase of evolution. 
Several epochs of mass exchange may punctuate the evolution of a binary until ultimately both components become white dwarfs, with masses and mass ratios very different from their initial values. 
Fig.\,\ref{fig_channels} provides a simplified illustration of how a series of binary star interactions can lead to formation of a close double-white dwarf binary system. 

Tracing the evolution of binary-star interactions as a function of initial masses and orbital separation ($m_1, m_2, a$) has been a major occupation of stellar theory for half a century \citep{Paczy1966,Webbink1976,Eggleton1983,iben84,Han1995,han98,Nelemans2000}.
As white dwarfs (WDs) represent the end state of almost all  stars,  double WDs are expected to be produced in many binary evolution channels. 
Among double WDs, about half  are expected to be double helium WDs. 
These usually form from systems in which both stars have masses less than 2.3\Msolar and which lose their envelope before core helium ignition. 
For instance, a pair of 1.4\Msolar and 1.1 \Msolar main-sequence stars in a 40-day orbit will produce a close binary system of two helium white dwarfs \citep{nelemans01}. 
In contrast,  double CO white dwarfs form in relatively wide and massive systems in which both stars can evolve to the AGB phase and leave a CO core after mass loss. 

Given the range of possible double white dwarf binaries that have emerged from these simulations, the potential for a radically new branch of stellar arithmetic was realised \citep{nomoto77,webbink84,iben84}.
Outcomes depend on whether one adds two helium white dwarfs (He+He), two carbon-oxygen white dwarfs (CO+CO), or one of each (He+CO), and also upon the mass ratio.

\subsection{Conditions for merger} 
\label{dwd_conds}

Once a DWD has formed, little happens for a long time.  
The binary acts as like a rotating dipole, its mass creating gravitational waves which remove energy from the system and cause the orbit to shrink. 
This spiral-in phase is interesting from the point of view of its gravitational-wave signature since unresolved DWDs are likely to provide the dominant foreground noise  at low frequency\citep{Webbink1998,Nelemans2001gw}. 
However, as contact is approached, the GW frequency and amplitude will rise more and more rapidly providing a `chirp' analogous to those observed in neutron-star and black-hole mergers \citep{gw170817,grb170817}, 
which will stand out form the foreground \citep{yu10,yu11,Kremer2017}. 

After a time given approximately as
\[ \tau \approx 10^7 (P/{\rm h})^{8/3} \mu^{-1} (M/\Msolar)^{-2/3} {\rm yr} \] 
\citep{marsh95b}
and 
the less massive white dwarf (subscript 2) will fill its Roche lobe and mass transfer will commence.
The equation shows that merger times for DWDs with periods $P > 6$\,h will exceed $10^{9}$yr, 
and a mass term with $\mu^{-1} (M/\Msolar)^{-2/3} < 1 $ will increase that time further\footnote{$M = m_1+m_2$, $\mu=\frac{m_1 m_2}{M}$}.  
 
Since the white dwarf mass-radius relation requires $r \propto m^{-1/3}$,  the donor will expand upon reaching contact,  when $P\approx3$\,m. 
To conserve angular momentum, the orbit will widen ($a$ will increase) if the mass ratio $q \equiv m_2/m_1 < 1$.
The donor expansion will exceed the Roche lobe expansion $\dot{r}_2>q\dot{a}$ if $q \gtrsim 2/3$, 
or less if mass transfer is non-conservative \citep{chen12}. 
Since the dynamical timescale is short and $\approx P_{\rm orb}$, the donor will totally disrupt
and form a hot shell around the accretor. 
If $q$ is less than some limit, mass transfer will be stable and have the characteristics of an AM\,CVn binary \citep{Nelemans2001amcvn,Marsh2004}. 
The location of the boundary between dynamical and stable mass transfer remains unclear and has some bearing on the types of outcome \citep{Motl2007,dan14}.

\subsection{Pre-merger candidates}

An early argument against DWD mergers as progenitors of Type Ia supernovae and other outcomes was a lack of DWDs. 
This  observational deficit was first corrected by \citet{marsh95b} and subsequently, through large-scale spectroscopy surveys including the Supernova Progenitor SurveY \citep[SPY:][]{napiwotzki01}, the Sloan Digital Sky Survey \citep[SDSS:][]{eisenstein06,breedt17} and the Extremely Low-Mass white dwarf survey \citep[ELM:][]{Kilic10a,Kilic11a,Kilic12a}, including many DWDs with mass ratios exceeding the critical value for merging and a substantial number which will merge with a Hubble time (10\,Gyr) \citep{Brown2016}. 

Notable amongst these is the eclipsing DWD SDSS J0651+2844 which, with $P=765$\,s and $q = 0.5$, shows  spiral-in  at a rate $\dot{P}=-8.95\pm0.11\times10^{-12} {\rm s\,s^{-1}}$ fully consistent with gravitational wave radiation \citep{hermes12b}. 
Other short-period DWDs which will reach contact within $10^6$\,yr\footnote{$\tau_{\rm C} \approx \frac{3}{8} \frac{P}{\dot{P}} $} and are also GW verification sources 
include DWDs 
SDSS J0935+4411 ($P=1188$\,s, $q>0.44$) \citep{kilic14},
SDSS J2322+0509 ($P=1201$\,s, $q=0.89$) \citep{Brown2020}, 
ZTF J1539+5027 ($P=415$\,s, $q\approx0.3$, $\dot{P}=-2.373\pm0.005 \times10^{-11}{\rm s\,s^{-1}}$ ) \citep{burdge19a} and
PTF J0533+0209 ($P=1234$\,s, $q=0.26$, $\dot{P}=-3.9\pm0.8 \times10^{-12}{\rm s\,s^{-1}}$ ) \citep{burdge19b}. 
Most appear to be CO+HE WDs, but J2322+0509 stands out for being an He+He WD.  

\subsection{DWD statistics}

By assuming that half of the star systems in the Galaxy are binaries and using the pioneering methods of binary-star population synthesis, \citet{han98} and \citet{nelemans01} obtained the first reliable statistical estimates of the galactic populations of double white dwarfs. 
Both \citet{han98} and \citet{nelemans01} found that the birth rate for close double white dwarfs in the Galaxy is $3.2\times10^{-2}\,\rm yr^{-1}$.
In more recent calculations, the {\it birth rate} for close double white dwarfs has been estimated as  $2.98\times10^{-2}\,\rm yr^{-1}$\citep{Liu2010wd} and $3.21\times10^{-2}\,\rm yr^{-1}$\citep{yu10}. 
Convolving with the star-formation rates and the evolutionary lifetimes, the total number of close binary white dwarfs currently present in the galactic disk is $\sim 3\times 10^8$ \citep{yu10}.
Of the double white dwarfs, $38\%$ are estimated to be He+He, $15\%$ are CO+CO, and $39\%$ are He+CO; the remainder  include an ONeMg WD \citep{yu10}.
Among these, the {\it merger rate} is estimated from $2.2\times10^{-2}\,\rm yr^{-1}$\citep{nelemans01} down to $2.7\times10^{-3}\,\rm yr^{-1}$\citep{yu10}. 
\citet[][Fig. 4]{yu11} separate the Galactic merger rates into He+He: $2\times10^{-3}\,\rm yr^{-1}$, CO+He: $6\times10^{-4}\,\rm yr^{-1}$, CO+CO:   $6\times10^{-4}\,\rm yr^{-1}$, and ONe+:  $3\times10^{-4}\,\rm yr^{-1}$.

It isn’t easy to estimate the merger rate form observation. 
\citet{brown20a} calculated the merger rate for low-mass double WD binaries from the ELM survey.
They infer that the latter are mostly He+CO WD binaries with a merger rate of $2\times10^{-3}\,\rm yr^{-1}$. 
The 1:1 number ratio of systems with merger times ($\tau$) more or less than 10 Gyr implies that the ELM WD merger rate exceeds the formation rate of AM\,CVn binaries in the Milky Way and hence supports models in which {\it most} He+CO WDs merge \citep{Shen2015}.

\begin{figure}
\centering\includegraphics[trim=30 30 0 0, clip, height=.27\textheight]{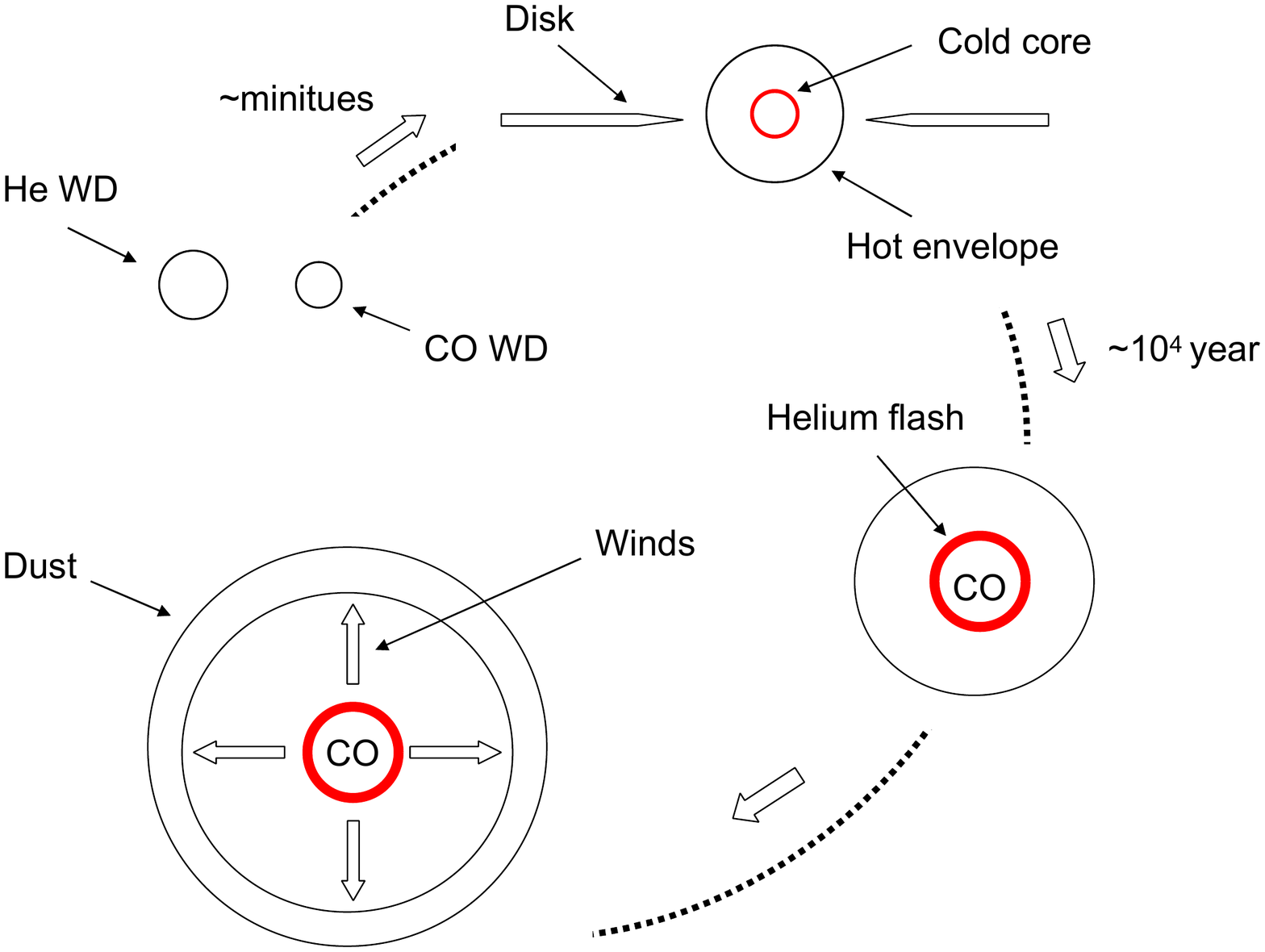}
\caption{Schematic of possible steps in a carbon-oxygen – helium white dwarf merger resulting in the formation an R\,CrB star (reproduced from \citet{zhang14})}
\label{fig_merger}
\end{figure}

\section{DWD merger physics}

Modelling the merger of two white dwarfs poses enormous computational challenges. 
Whilst  we probably understand what the major phases look like, but we are also probably a long way from knowing precisely what the temperature, density and angular momentum distribution is at any stage of process.

First attempts were based on a three-dimensional hydrostatic code \citep{hachisu86a,hachisu86b} and provided hints that, even for $M_{\rm Chandrasekhar}<M<2.4\Msolar$ in a CO+CO merger, the maximum density would be too low to ignite carbon deflagration, unless angular momentum is lost during the merger. 
It was already known that, if thick disk physics applies, there is little to stop such angular momentum loss \citep{pringle75}. 
Consequently, a WD merger is primarily a hydrodynamics problem covering many pressure and density scale heights, with  high-density and nuclear physics and ultra-short time scales to spice things up. 

A summary of three decades of progress provides a thumbnail sketch of a DWD merger (Fig.\,\ref{fig_merger}).

Subsequent calculations adopted the smoothed-particle hydrodynamics (SPH) approximation 
to demonstrate how material starts to be stripped from the donor WD and spiral down to the accretor's surface, followed quickly by complete tidal disruption of the donor. 
\citep{benz90,segretain97,guerrero04}.
As the mass transfer rate accelerates, matter cannot be assimilated directly by the accretor, so starts to accumulate in a high-entropy high angular-momentum shell around the more massive white dwarf. 
The next generation of models provided the insight that the merger debris can be described by a Keplerian disk-like shell lying outside a rapidly-rotating, high-entropy `hot' envelope. 
The relatively undisturbed accretor behaves as a slowly rotating 'cold' compact core \citep{yoon07,loren09}. 
Interesting nucleosynthesis may occur at the base of the 'hot' envelope, where protons, $\alpha$ particles and light elements mix and can briefly reach temperatures in excess $10^9$ K \citep{clayton05,longland11,longland12,menon13,menon19}. 

More recent models adopt the full hydrodynamics approach, including some nucleosynthesis.
\citet{staff12} performed a series of hydrodynamic simulations of merging double WD systems for RCB stars.
By assuming a thick helium buffer layer on the primary WD surface, they found significant production of $^{18}O$.
Thus, the ratio of $^{18}O/^{16}O$ intensely depends on primary WDs. 
This result is also confirmed by a recent simulation\citep{staff18}. 
By adopting the outcome of  \citet{staff12}, \citet{menon13,menon19} represent the surface abundances of RCB stars with a particular mixing profile. 

Calculating the merger of two unequal mass CO WDs with viscosity, \citet{shen12} argues that the remnants will take a relatively long-time evolution during the envelope heating phase and then yield a high-mass ONeMg WD or NS rather than SN Ia,  reflecting the conclusion of \citet{hachisu86a}.  
These results indicate that the detail of the physical model plays an essential role in the fate of double WD mergers.
Hence, \citet{Shen2015} considered viscosity in the double WD mergers with more extreme mass ratios.
The outcome for such systems, which was previously thought to be dynamically stable mass transfer, is more likely to be a merger. 
In the low mass (or sub-explosion) domain, this will result in a higher ratio of merger products to 
mass-transferring AM\,CVn systems, essentially as observed \citep{brown20a}.

Despite progress in the dynamics and nucleosynthesis of the few minutes surrounding the merger itself, questions persist  about the subsequent evolution. 
How quickly does the high-entropy envelope `accrete' onto the primary -- is this fast, or slow, or some combination? 
How deeply are the surface layers of the accretor disturbed, and how completely are these layers and any nuclear products formed in the high-T flame, mixed through the surface layers of the product?
How soon after the merger is quasi-stable nuclear burning established? 
And, of course, what are the conditions that can trigger a runaway explosion?

\begin{figure}
\centering\includegraphics[trim=10 10 0 0, clip, width=.48\textwidth]{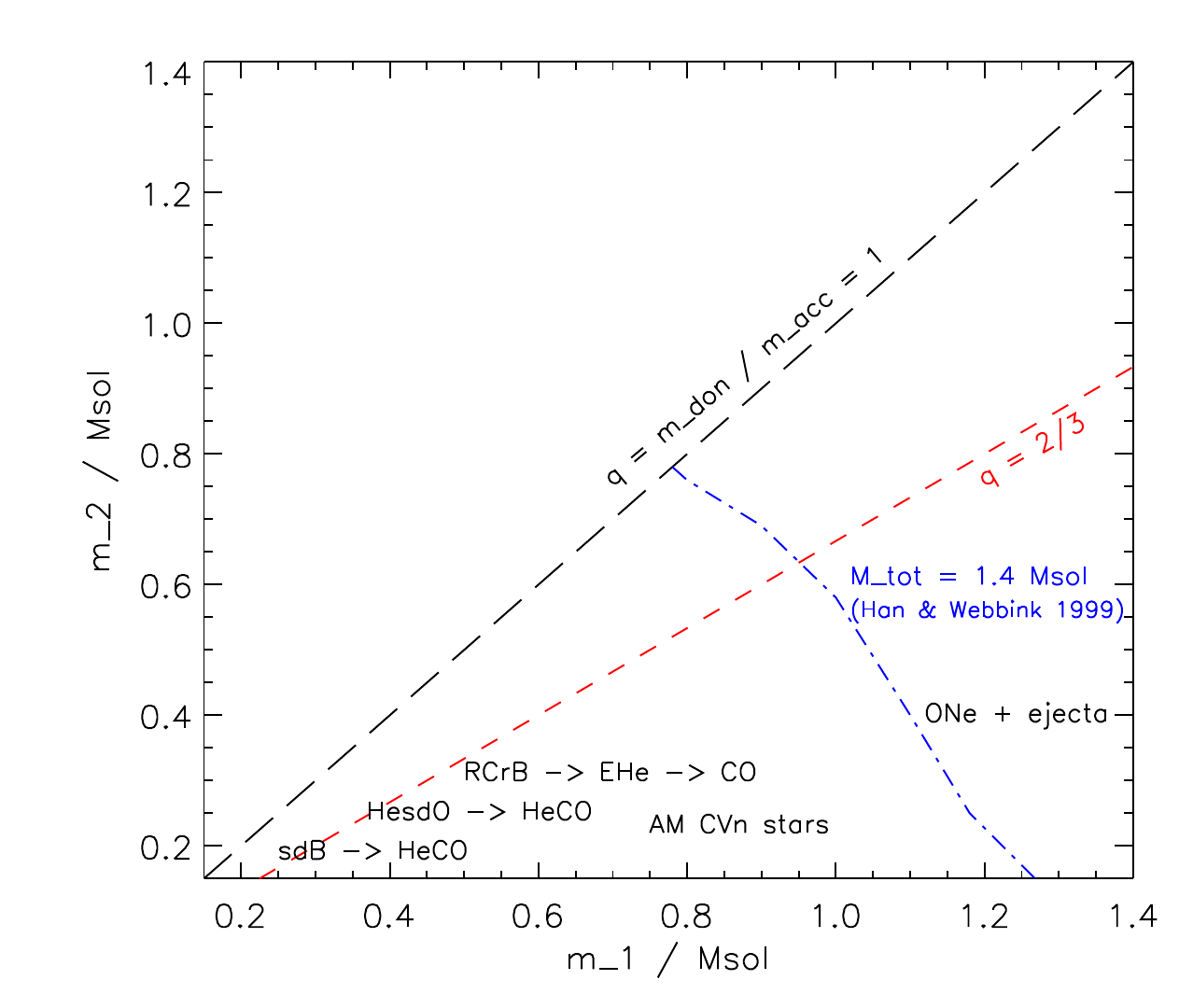}
\caption{Schematic distribution of surviving DWD merger products as a function of accretor and donor mass (adapted from \citet{dan14}). See the latter for explosions. }
\label{fig_products}
\end{figure}

\section{DWD merger products}

Key to the outcome and subsequent evolution of the merger is the physics of the interaction itself. However, one thing is clear; since neither participant contains much hydrogen, and what hydrogen has survived will be mixed with substantially more helium and possibly other material, the product will be  hydrogen-deficient.

\subsection{DWD arithmetic} 
\label{dwd_arith}

Schematically, DWD merger arithmetic has been articulated for nearly 40 years. 
First steps to compute post-merger evolution used quasi-static 1-dimensional models to approximate the merger. Three general cases are identified below, together with a non-exhaustive list of the major calculations:\\[1mm] 
\noindent $\bullet$
{\bf He+He} $\rightarrow$ He shell ignition \\
$\rightarrow$  He core burning (hot subdwarf) $\rightarrow$  He/CO WD \\[1mm]
\citep{nomoto77,nomoto87,kawai87,kawai88,iben90,saio00,zhang12b,zhang12a,hall16,schwab18}\\[1mm]
\noindent $\bullet$
{\bf He+CO} $\rightarrow$ He shell ignition $\rightarrow$ explosion $|$ RCB star \\
$\rightarrow$ EHe star $\rightarrow$ CO WD  \\[1mm]
\citep{webbink84,iben84,iben90,saio02,zhang14,staff18,lauer19,schwab19,crawford20}\\[1mm]
\noindent $\bullet$
{\bf CO+CO} $\rightarrow$ C shell ignition $\rightarrow$ C core burning \\
$\rightarrow$ ONeMg WD  $|$ explosion \\[1mm]
\citep{hachisu86a,kawai87,nomoto87,Mochkovitch1990,segretain97,saio98}\\[1mm]
Additional permutations are possible if we distinguish hybrid He/CO WDs. 

\subsection{Post-merger evolution}

The simplest case, the merger of two helium white dwarfs to produce a core helium-burning star has latterly been a key theme in the study of hot subdwarfs \citep{han02,zhang12a,heber16}. 
Whilst He+CO and CO+CO WD mergers provide a major channel for the production of Type Ia supernovae,  dramatic events with huge significance for cosmology, we are more interested here in stars which survive the merger. 
These are primarily considered to include the helium-rich subdwarfs, extreme helium (EHe) stars and R\,CrB (RCB) variables.

Whilst all three DWD cases involve post-merger ignition of fresh fuel in a shell around a degenerate core, the first and third are notable because the enclosed core of previously spent fuel re-enters the energy budget of the star. Working inwards from the initial interface between accretor and accreted material, core degeneracy is progressively lifted, fresh fuel is ignited, until central core burning is established.

The handover from the violent hydrodynamical event to subsequent quasi-stable evolution has been treated in a number of ways. 
Early calculations \citep{saio00,saio02,saio98}  assumed relatively 'slow' accretion $(\sim 10^{-5}\Msolar {\rm yr^{-1}})$ of hydrogen-deficient material onto a model of an evolved white dwarf. 
In this case, too much hydrogen in the mix would precipitate a nova-like explosion. 
With improved observations of the surface chemistry of proposed merger products, 
\citet{zhang12a,zhang12b,zhang14} used outputs from hydro calculations \citep[e.g.][]{loren09} to indicate the rate of accretion and the chemistry of the accreted material.
Again, the quasi-static approximation was adopted even for 'fast' accretion rates of up to $10^4\Msolar {\rm yr^{-1}}$ to represent a 'hot' envelope
 which is produced in hydro simulations.  
 
More recent calculations have been carried out by \citet{schwab18} for He+He mergers and by \citet{schwab19,lauer19} and \citet{crawford20} for He+CO mergers. 
These employ recent versions of the stellar evolution toolkit {\sc mesa} \citep{paxton11,paxton13,paxton15,paxton18} and povide more detailed accounts of surface compositions of merger products as functions of initial metallicity, the nuclear networks included and the degree of mixing encountered during and after the merger. 
Apparent from these studies and well-known from previous work on asymptotic-giant branch stars, thin-shell helium burning is numerically  difficult to model; mesh points have to be finely distributed and timesteps have to be short for any chance of success.  
However, a common result is that the surfaces of the merger products are produced by mixing during or immediately after the merger, and hence the temperature of the helium-burning shell at ignition has a strong influence on the yields of both CNO-process and $\alpha$-capture isotopes.  

From the He+He models of  \citet{hall16} and \citet{schwab18}, we may expect to observe that helium-rich hot subdwarfs lack hydrogen, with $n_{\rm H}<10^{-4}$ by number. 
Observationally,  $n_{\rm H}<10^{-2}$ might be more realistic \citep{naslim11,stroeer07}. 
We also expect the helium-rich hot subdwarfs to be separated into carbon-rich and nitrogen-rich subgroups by composition and surface effective temperature. 
This is related to their masses, as suggested by \citet{zhang12a}. 

For CO+He models a critical feature is the ratio of isotopes, which indicate the degree of burning during the merger. 
The ratios of \iso{12}{C}/\iso{13}{C} and \iso{18}{O}/\iso{16}{O} have been measured in some RCB stars \citep{clayton07,garcia10,hema12}. 
\citet{schwab19} argues that the merger model has a thermal reconfiguration phase, which could last for 1 kyr. 
A test would be to identify such RCB precursors, cool H-deficient stars with lower luminosities, secular brightening, and in apparent dusty shells.

\subsection{Post-merger candidates}

\S\S\,\ref{dwd_conds} and  \ref{dwd_arith} introduced conditions for a DWD merger and the identification of potential merger products respectively. 
More complete treatments of orbital stability and the conditions for detonation have been given by a number of authors \citep[e.g.][]{han99,nelemans01,chen12,dan14}.
Fig.\,1 in \citet{dan14} provides a compelling illustration of merger outcomes as a function of accretor mass and donor mass ($m_1$ and $m_2$ above). 
The approximate loci of post-merger survivors are reproduced in Fig.\,\ref{fig_products}, 
where the post-merger mass  $M = M_{\rm Chandrasekhar} \approx 1.4\Msolar$  has been given assuming non-conservative mass transfer \citep{han99} and the mass ratio $q=2/3$, commonly used a stability criterion in early work, has also been added. 
 
\begin{figure*}
\centering\includegraphics[trim=15 25 0 0, clip, width=.85\textwidth]{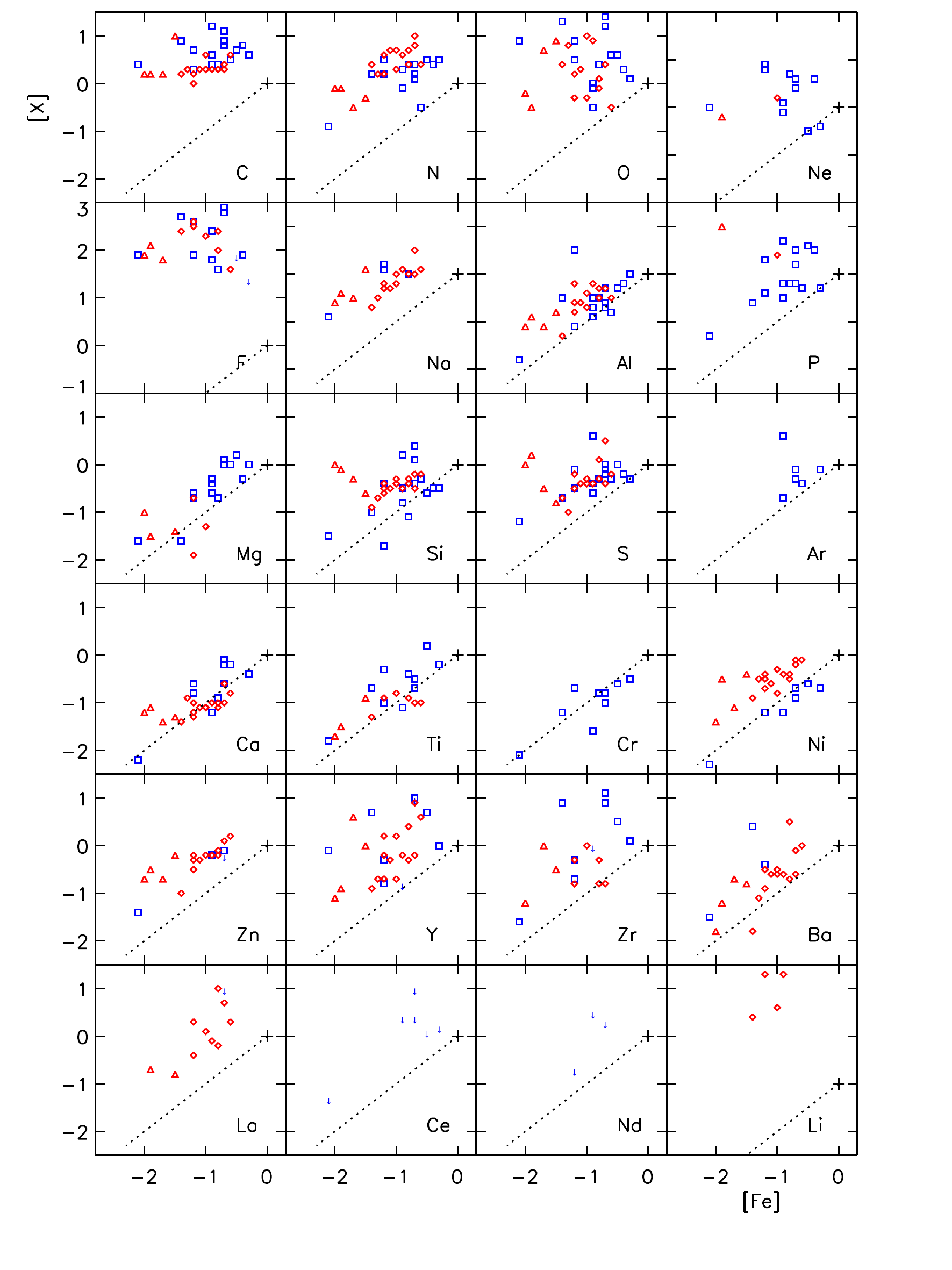}
\caption{Observed surface abundances (log number relative to solar) versus iron abundance (same units) for EHe stars (blue squares), majority RCB stars (red diamonds) and minority RCB stars (red triangles). Upper limits are shown as arrows. The dotted line indicates a solar composition scaled to iron. Note that the plots for Ne, F and Li are offset vertically; a cross indicates $[\rm{Fe}], [X] = 0, 0$ (solar abundance) in every case. Adapted from \citet{jeffery11a} with updated values from \citet{kupfer17} and \citet{bhowmick20}.  }
\label{fig_abunds}
\end{figure*}

\section{Observational tests: He+CO models}

Primary tests of merger models are to match observed masses, luminosities, and radii. 
None of these are in general directly observable, although Gaia has recently started to address the question of luminosities for some putative merger products \citep{martin17a} (Martin \& Jeffery in prep.).
Since effective temperatures and surface gravities can be measured directly, these provide a proxy test for overall dimensions.
In  cases where stars are pulsating, distances have been estimated independently using the Baade method \citep{jeffery01c,jeffery01b,woolf02c}. 
Pulsation properties, including period $P$ and period changes $\dot{P}$ also give estimates for mean density and thermal timescale, both of which can be compared with evolution models,
with remarkable success in the case of V652\,Her \citep{saio00}. 
The surface chemistry of DWD merger products provides an additional test which is not available for classes of merger, such as blue stragglers, which leave a hydrogen-rich surface.  
In this paper, we therefore focus on the surface chemistry tests. 

Merger products almost uniquely expose material which has previously been at the center of a star.
Hence we expect to see either pp- or CNO-processed helium from the donor, 
as well as other nuclear ash which may come either from the outer layers of the accretor or from the merger process itself. 
If the accretor is a CO WD, the ash includes material from the intershell region of the AGB star progenitor, and carbon-enriched material from the outer layers of the accretor's  core.
Carbon may also come from the donor if it is sufficiently massive to be an He/CO white dwarf. 


\begin{table}
\caption{Abundance trends in RCB and EHe stars.}\label{tab_abunds} 
    \centering
    \begin{tabular}{ll}
    \hline
Observed  & Process \\ 
\hline
H $\lesssim 10^{-3}$ 	&  relic \\
Ca,Ti,Cr,Mn,(Ni)  $\propto$ Fe    &	primordial \\ 
N/Fe   $\propto$ [(C+N+O)/Fe]	 & CNO \\
C/Fe $\gg 0$, $\iso{12}{C} \gg \iso{13}{C}$  &	$3\alpha$   \\
O/Fe $\gg 0$, $\iso{18}{O} \approx \iso{16}{O}$   \\
Ne/Fe $\gg 0$	& $\iso{14}{N}+2\alpha$ \\
Mg,Si,S,…  	&	X $+\alpha$  \\ 
F/Fe $\gg 0$ &	p- or n-capture \\
light s/Fe $\gg 0$ &  ??   \\
heavy s/Fe  $\propto$ Fe + c &		AGB ?   \\
P/Fe $\propto$ Fe + c &	AGB ?: $m_{\rm AGB}\approx 2 - 3\Msolar $  \\
Li present	&	\iso{3}{He} + p \\
\hline
    \end{tabular}
\end{table} 

\subsection{Understanding EHe and RCB abundances}

\citet{jeffery11a} compiled and standardized surface abundance measurements of 18 EHe and 
18 RCB stars from papers by \citet{jeffery88,jeffery93a,jeffery93b,jeffery98,drilling98,asplund00,pandey08,pandey01,pandey06a,pandey06b,pandey06c} and \citet{pandey11}.
Subsequent work on fluorine \citep{bhowmick20} and on BD$+10^{\circ}2179$ \citep{kupfer17} is included in Fig.\,\ref{fig_abunds}\footnote{ Recent work suggests  DY\,Cen  should not be included and is probably a  late  shell-flash post-AGB star \citep{jeffery20a}.}.  

Early work recognised that the surfaces of EHe and RCB stars comprise at least three components, namely (i) a trace of primordial `stuff' represented by hydrogen, (ii) a predominance of CNO-processed material represented by the high helium abundance (97\%) and a super-solar nitrogen abundance, and (iii) excess carbon from 3-$\alpha$ processed helium \citep{schoenberner86a}. 

To a large extent and up to around 2007, abundance studies aimed to resolve the origin of EHes and RCBs in terms of competing evolutionary models. 
As well as the DWD merger (DD) model \citep{webbink84}, a single-star model involving a late thermal pulse or final shell flash (FF) in a contracting post-AGB WD was considered \citep{schoenberner79,iben84}. 
Prior to 1984, several other {\it ad hoc} models had been circulating \citep[e.g.][]{paczynski71,schoenberner77}. 
The DD or FF question was finally resolved by the discovery of very high abundances of \iso{18}{O} in several RCB stars \citep{clayton07,garcia10}, prompting \citet{clayton07} to reach the same conclusion as \citet{saio02} that the DD origin is more likely.  

\citet{jeffery11a} extends the broad decoding of the EHe and RCB surface abundances in terms of correlations described in Table\,\ref{tab_abunds} and Fig.\,\ref{fig_abunds}.  
What additional clues about previous evolution can be inferred?

Where the abundances of iron and other elements unaffected by light-element nucleosynthesis (e.g. calcium, titanium , chrmomium, manganese, nickel) can be measured directly, their abundances are correlated and indicate overall metallicity at formation. 
 
Nitrogen scales with iron; more specifically, nitrogen corresponds to the expected sum of carbon, nitrogen and oxygen, assuming that  these scale with the iron abundance at formation.
Hence nitrogen may be taken as a proxy for metallicity at formation {\it and} provides evidence that the helium has been produced from the hydrogen-burning CNO cycle, in which most of the carbon and oxygen is converted to nitrogen, with \iso{14}{N} being the longest-lived isotope in the cycle.  

If the material was entirely CNO-processed helium, carbon and oxygen should be depleted. 
This is true for V652\,Her and some helium-rich subdwarfs \citep{jeffery99}, but not the majority of EHe and RCB stars. 
Instead, carbon is generally 2 -- 10 times solar, independent of metallicity. 
Where isotopes can be measured, \iso{12}{C} is much more abundant than \iso{13}{C} \citep{hema12}. 
Both point to the presence of 3-$\alpha$ processed helium. 

Whilst not consistently super-solar, oxygen shows a wide range of enhancements by 1 to 3 dex.
This was a puzzle up to the discovery that \iso{18}{O} can be as abundant as \iso{16}{O} \citep{clayton05}, indicating that excess oxygen can be attributed to $\alpha$ captures on the already overabundant \iso{14}{N}. 
Normally, \iso{18}{O} is destroyed promptly by an additional $\alpha$ capture to \iso{22}{Ne}; it only survives if the $\alpha$ exposure is short. 
So where is \iso{18}{O} made? 
In some models, the outer layers of a CO WD contain a \iso{18}{O} pocket, arising from incomplete conversion to \iso{22}{Ne} during the transition from AGB to WD. 
In a `hot' merger, \iso{18}{O} can be produced by heating \iso{14}{N} from the He WD. 
\citet{jeffery11a} argue that either can account for the oxygen excess in EHe and RCB stars. 

Neon is almost universally overabundant by 1 - 2 dex, and can be accounted for by two  $\alpha$ captures on \iso{14}{N}. 
Again, the source must be linked to the source of \iso{18}{O}, the latter being a residue. 

Sodium, aluminium, magnesium, silicon and sulphur are broadly correlated with metallicity, but with varying evidence of enhancement compared with, say, calcium, titanium, chromium and nickel. 
Three  groups of elements are of particular interest: fluorine, phosphorus and s-process elements (yttrium, zirconium, barium, \ldots).

Fluorine is ubiquitously 2 -- 3 dex supersolar, independent of metallicity.
Some fluorine may  be produced by proton captures in the helium intershell during the evolution of 2 -- 3 \Msolar stars on the AGB, but not in sufficient quantity to match observation \citep{jeffery11a}.
\citet{zhang12b} suggest that fluorine arises from the reaction 
$\iso{14}{N}(\alpha,\gamma)\iso{18}{F}(p,\alpha)\iso{15}{O}(\alpha,\gamma)\iso{19}{Ne}(\beta^{+})\iso{19}{F}$ during a hot merger, but also underpredict the observed abundances.
\citet{menon13} note that another source is the He-burning shell of the post-merger star, with 
$\iso{13}{C}(\alpha,n)\iso{16}{O}$ providing neutrons to seed the reaction 
$\iso{14}{N}(n,p)\iso{14}{C}(p,\gamma)\iso{15}{N}(\alpha,\gamma)\iso{19}{F}$. 
\citet{bhowmick20} argue that this provides a more plausible route. 
  
The origin of phosphorous, which is mostly 1 -- 2 dex enhanced and remarked on by \citet{kaufmann77}, is less clear. 
Like fluorine it is produced in the helium intershell of 2 -- 3 \Msolar\ stars on the AGB. 
The fact that fluorine is universally 2 -- 3 dex supersolar, whilst phosphorous generally scales with metallicity suggest that their origins in EHe and RCB stars are different.  
By analogy with the carbon and nitrogen abundances, we suggest that the metallicity-independent excess of fluorine is likely attributable to fusion during a hot merger, and the more scaled behaviour of phosphorous to production in the AGB precursor. 
A contrary argument would benefit from additional phosphorous measurements from RCB stars, especially the metal-poor minority group. 

Large overabundances of the light s-process elements yttrium and zirconium are seen in both EHe and RCB stars with excesses of 0 -- 2 dex distributed randomly with metallicity.   
The heavy s-process elements barium and lanthanum are also observed but are generally only 0 -- 0.5 dex (Ba) and 0.5 -- 1 dex (La) above the scaled value. 
Again, the difference in distribution suggests that the heavy s-process elements were formed in the AGB precursor, whilst the light s-process elements are formed during the merger. 

In four cases, lithium is observed with a 3 -- 4 dex overabundance. 
Normally destroyed by proton-captures at moderate temperatures, \citet{longland12} and
\citet{zhang14} argue that it could form during a hot merger via the $\iso{3}{He}(\alpha,\nu)\iso{7}{Li}$ reaction using \iso{3}{He} from the He WD. 
This could provide sensitive test of the merger models, which currently predict a deep but short convective phase to bring `hot-merger' products to the post-merger surface. 
Under what conditions would fresh lithium survive or be destroyed?  

As already remarked with reference to fluorine, phosphorous, and the s-process elements, an intriguing  feature of Fig.\,\ref{fig_products} is that some elements scale as iron, others scale roughly as iron plus a constant, and others are enhanced independent of metallicity. 
The latter group includes carbon, oxygen, neon, fluorine, yttrium and zirconium. 
It would appear that these elements are produced during the merger, and that amount of product is  sensitive to the specific conditions of the merger rather than the initial composition. 
It will be interesting to see if these yields correspond with some other property of the post-merger star, such as its mass or luminosity.
Establishing correlations such as those between fluorine and other elements explored by \citet{bhowmick20} will provide stringent tests of the merger models. 

\subsection{The simple recipe}

Building on `a few mere facts' \citep{lambert94,rao96} concerning the surface abundances of RCB stars,  \citet{lambert96} proposed `a simple recipe' in which processes that alter chemical composition $A$ between evolutionary stages $i$ and $j$ are represented by an operator $O_A$  such that 
\[ A_j = O_A(i,j) A_i. \] 
Inspired by this insight, one of us attempted to build a simple algebraic model that included the proportions of primordial, helium-rich and carbon-rich material as unknowns, and operators that reflected known properties of the CNO cycle and 3$\alpha$ processes.   
The goal was to use observed surface abundances to make inferences about the progenitor processes. 

The discovery of enhanced s-process elements in several RCB  stars \citep{bond79,lambert94} led to a recognition that the surface of a He+CO WD merger would also require knowledge of both the hydrogen distribution in the He WD and of the chemical composition of the helium-enriched intershell region of the CO WD. 
Consequently, the `simple recipe'  adopted by \citet{jeffery11a} considered the contributions of five layers of material originating from two stars, each having a representative composition defined by contemporary stellar evolution theory, the latter providing the operators $O_A$. 

Meanwhile, the discovery of very high abundances of \iso{18}{O} in some RCB stars led to experimental calculations of fast nucleosynthesis in a hot shell \citep{clayton07} and the conclusion that nucleosynthesis during the merger strongly affects the post-merger  surface composition. 
This an additional operator was introduced, taking a different form according to whether a `hot' merger or a `cold' merger had occurred. 

The `simple recipe' models provided evidence that EHe/RCB progenitors include an AGB star with inital mass in the range 1.9 -- 3 \Msolar, but could not resolve the source of all observed abundance anomalies at the time \citep{jeffery11a}.  
Whilst simple in comparison with full {\sc mesa} calculations, for example, the model provides rapid estimates of merger outcomes for a wide range of parameters, including progenitor masses and metallicities, and hence allow various scenarios to be tested quickly. 
Further exploration which combines recent hydro- and {\sc mesa-} outputs with the simple recipe would be instructive. 

\begin{figure*}
\centering\includegraphics[trim=0 0 0 0, clip, width=.85\textwidth]{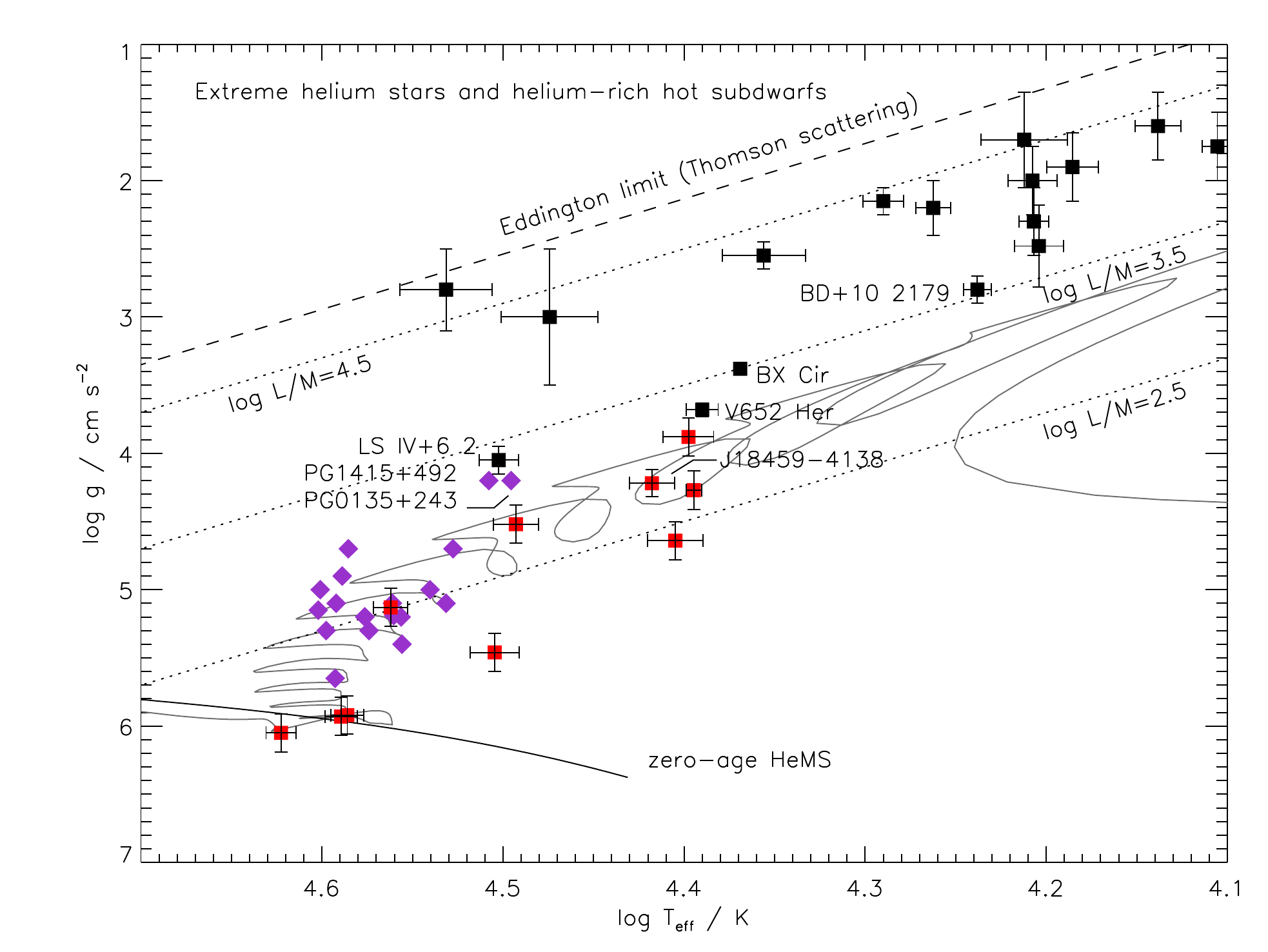}
\caption{{\it Preliminary} surface properties of helium-rich stars observed in the SALT survey (red squares, Jeffery et al. in prep.) compared with EHe stars (black squares), extreme helium subdwarfs (magenta diamonds). 
Stars mentioned in the text are labelled.
The positions of the Eddington limit (Thomson scattering: dashed), luminosity-to-mass contours (solar units: dotted) and zero-age helium main-sequence (HeMS) are also shown.
The post-merger evolution track for a model of a  He+He white dwarf merger \citep[0.30+0.25\,$M_\odot$]{zhang12a} is shown.
}
\label{fig_salt}
\end{figure*}

\section{Observational tests: He+He models}

Post-merger candidates for He+He WD mergers are primarily the high-gravity helium-rich subdwarfs; being core helium-burning stars their lifetimes are $\sim10^8$\,y and hence are the most numerous.    
Additional helium-rich stars with surface gravities intermediate to the subdwarfs and EHe stars discussed above may be in a more immediate post-merger phase, having a helium-burning shell that is burning inwards.
The best example is V652\,Her \citep{jeffery99b}. 
To date, information on the surface chemistry of both groups is limited. 
Most candidates show nitrogen-enhanced surfaces, evidence of a CNO-processed surface. 
Several also show carbon-enhancement; \citet{zhang12a} demonstrate that carbon produced by $3-\alpha$ reactions during or immediately after the merger is mixed to the surface, providing the subdwarf is sufficiently massive. 

At present, this roughly represents the limit of our knowledge. 
Abundances of iron, other iron-group metals and s-process elements are difficult to measure in hot subdwarfs observed at optical wavelengths; and there are few high quality ultraviolet observations of extremely helium-rich hot subdwarfs. 
Cooler candidates, like V652\,Her, are extremely rare. 
The status of the carbon-rich BX\,Cir is unclear.
Its mass and luminosity are similar to V652\,Her, but is it a post CO+He merger or a post He+He merger? 

Unlike the case with the hot EHe stars and cool RCB stars, there are no known cool hydrogen-deficient stars which might be the immediate precursors of V652\,Her. 
One reason is that H-deficient stars are hard to recognise at temperatures $\teff<10\,000$\,K where helium absorption lines cannot be observed.
A second reason might be that evolutionary lifetimes are extremely short; unlike the RCB stars which have a relatively long stable helium-shell burning phase, the shell in a post He+He merger moves inwards on a timescale of $\sim10^4$\,y. 
With an optimistic birth-rate of $\sim10^{-2} {\rm Galaxy^{-1} y^{-1}} $, this would give only $\sim 100$ such stars in the Galaxy.  

In an attempt to address the shortage of known helium stars on the post-merger track, one of us commenced a spectroscopic survey of chemically-peculiar subdwarfs, targeting all southern hot subdwarfs classified as helium-rich \citep{jeffery17c},(Jeffery et al., in prep.). 
  An early discovery was GALEX J$184559.8-413827$, \citep{jeffery17b}, which is almost a twin of V652\,Her; it is also nitrogen-rich but not yet detected to pulsate. 
Analyses of several others are currently in progress.  
Preliminary results for a subsample from the survey are shown on the $\teff-g$ plane in Fig.~\ref{fig_salt}.

Emerging from this diagram, which does {\it not} represent a complete sample, is a sequence of extremely hydrogen-deficient stars running approximately along the predicted loci of post-merger models He+He WD mergers \citep{zhang12a}. 
 
Another feature emerging from this figure is the proximity of the carbon-rich `EHe' star LS\,IV$+6^{\circ}2$ to two helium-rich subdwarfs PG\,1415+492 and PG\,0135+243. 
The majority of EHe stars have much higher luminosity-to-mass ratios ($L/M$) but a few, including BD$+10{^\circ}2179$, have $\log\,(L/\Lsolar)/(M/\Msolar) \approx 3.5$. 
Since post-merger models predict that stars will contract at roughly constant luminosity, 
discerning which stars, if any, are post-He+He or post-CO+He mergers may not be easy.

\section{Conclusion}

We have presented a condensed summary of current science of double white dwarf mergers, at least as far as this pertains to the production of remnants that survive and are detectable as hydrogen-deficient giants and subdwarfs.  
The key points are that:
\begin{itemize}
\item Many Galactic double white dwarf binaries are evolving towards a dynamical merge. 
\item Some of these DWDs should be resolvable with a suitably sensitive GW detector. 
\item Merging events are intrinsically rare ( $< 1$ / century per galaxy for DWDs); they need to be distinguished from other events (e.g. late shell flashes).
\item Merging events and post-merger evolution can be simulated, but it is difficult and many challenges remain (hydrodynamics, nucleosynthesis, mixing).
\item Post-merger candidates can be identified from spectral and pulsation signatures.  Verification is harder, model-sensitive, but do-able from chemistry, asteroseismology, parallaxes, kinematics and models.
\item For DWD post-mergers: extremely helium-rich low-mass supergiants (including some R\,CrB stars) and extremely helium-rich hot subdwarfs look like very good bets.  
\item Most Galactic EHe supergiants have probably been found. Many new extreme helium subdwarfs and RCBs are still to be found.  
\end{itemize}

\section*{Acknowledgements}
The lead author is indebted to generations of collaborators on the topic of this paper. 
Particular thanks are due to Amanda Karakas and Hideyuki Saio, who cowrote \citet{jeffery11a}, Shenghua Yu, who did the population synthesis, Kameswara Rao, Gajendra Pandey and David Lambert, who have kept him grounded in observations, and Dick Carson, Phil Hill, Kurt Hunger, Tony Lynas-Gray, John Drilling and  Uli Heber, who set him on this fantastic journey of discovery.  
He is  indebted to the organizers of `150 Years of the Periodic Table' for their invitation and hospitality in Bengaluru, and to the organizers of the Tucson workshop on 'Stellar Remnants' in 1984 at which he first encountered Webbink's merger model for RCB stars \citep{webbink84}.  

\bibliographystyle{mnras}
\bibliography{ehe}



\label{lastpage}
\end{document}

%% file: mydefs.tex
\newcommand{\Msolar}{\mbox{\,$\rm M_{\odot}$}}        
\newcommand{\Lsolar}{\mbox{\,$\rm L_{\odot}$}}        

 \newcommand{\teff}{\mbox{\,$T_{\rm eff}$}}      

  \newcommand{\iso}[2]{\mbox{$^{#1}{\rm #2}$}}           
  \def\simge{\mathrel{\raise1.16pt\hbox{$>$}\kern-7.0pt
    \lower3.06pt\hbox{{$\scriptstyle \sim$}}}}           
  \def\simle{\mathrel{\raise1.16pt\hbox{$<$}\kern-7.0pt
    \lower3.06pt\hbox{{$\scriptstyle \sim$}}}}           